# Half a billion simulations: Evolutionary algorithms and distributed computing for calibrating the SimpopLocal geographical model

*Clara Schmitt*[1,2,3], *Sébastien Rey-Coyrehourcq*[1,2,3], *Romain Reuillon*[2,3], *Denise Pumain*[1,2,3,4]

[1] Université Paris 1 Panthéon -Sorbonne, 12 place du Panthéon, 75005 Paris, France

[2] UMR Géographie-Cités, 13 rue du Four 75006 Paris, France

[3] Institut des Systèmes Complexes Paris-Ile-de-France, 113 rue Nationale, 75013 Paris, France

[4] Institut Universitaire de France, 103 bd Saint-Michel, 75005 Paris, France

**Abstract**
Multi-agent geographical models integrate very large numbers of spatial interactions. In order to validate those models large amount of computing is necessary for their simulation and calibration. Here a new data processing chain including an automated calibration procedure is experimented on a computational grid using evolutionary algorithms. This is applied for the first time to a geographical model designed to simulate the evolution of an early urban settlement system. The method enables us to reduce the computing time and provides robust results. Using this method, we identify several parameter settings that minimise three objective functions that quantify how closely the model results match a reference pattern. As the values of each parameter in different settings are very close, this estimation considerably reduces the initial possible domain of variation of the parameters. The model is thus a useful tool for further multiple applications on empirical historical situations.

**Acknowledgments**
This work is funded by the ERC Adv Grant Geodivercity, the *Agence de l'Environnement et la Maitrise de l'Energie* (ADEME), the network *Réseau de Recherche sur le Développement Soutenable* (R²DS), the *Institut des Systèmes Complexes Paris Ile-de-France* (ISC-PIF). Results obtained in this paper were computed on the biomed and the vo.complex-system.eu virtual organization of the European Grid Infrastructure (http://www.egi.eu). We thank the European Grid Infrastructure and its supporting National Grid Initiatives (France-Grilles in particular) for providing the technical support and infrastructure. We also thank an anonymous reviewer for his/her help in improving the first version of the paper.

# The challenge of calibrating geographical multi-agent models

Geographical simulation models of systems of cities are based on the assumption that the microgeographical interactions are likely to generate the emergence of stylized dynamics on the macrogeographical scale, which constitute one of the recurring universal characteristics of these complex systems (Pumain and Sanders, 2013). The heterogeneity and the large number of possible scenarii of implementation to represent these processes, the individual-based description of dynamics, the non-linearity of interactions, the diversity of forms of spatial relationships and the importance of the historical context are so many reasons for geographers to regularly use agent-based modelling as a support for reflection and experimentation (Batty, 2008; Crooks et al, 2008; Heppenstall et al, 2011; Sanders, 2007). We introduce here a simplified model called SimpopLocal (Rey-Coyrehourcq, 2014; Schmitt, 2014) that extends the Simpop family of models (Pumain, 2009; 2011). It describes the emergence of a system of settlements where the process of innovation that generates the growth dynamics of the settlements is made endogenous.

What is at stake in a successful agent-based model in geography is to assemble a set of individual-based mechanisms adapted to the level of resolution of the problem and to evaluate its ability to answer this problem with a high level of confidence (Sargent, 2005). This evaluation usually requires a calibration stage (Balci, 1998) during which the ability of the model to reproduce a specific dynamic or structure is assessed. This calibration phase is generally conducted by a trial and error method. When applied on SIMPOP2 it resulted in being fairly satisfied with the calibration obtained thanks to a hundred simulations (Bretagnolle and Pumain, 2010). It was thus not possible to determine if the corresponding values found for the parameters, although tedious to estimate through this manual calibration procedure (each new increment in the parameter values can disturb the dynamics), were the best possible estimation or simply denoted the existence of a local minimum in the phase space of the dynamic model. It is thus crucial to automate the process of calibration and to substitute temporarily the human expertise of the model by an automated expertise, which requires a quantitative transposition of the expert evaluation.

To carry out this automation, we developed and implemented methods and tools within the community of research on the complex systems (Bourgine et al, 2009) and more precisely in the modelling platform called SimProcess (Rey-Coyrehourcq, 2013). After having described the geographical model Simpoplocal on which the method is tested, we present an automated procedure of calibration that allows a massive exploration of the value space of the parameters, which was up to now impossible to handle in the usual way. This exploration is guided by specific and relevant objectives as for the issues dealt with by SimpopLocal. After extraction and study of the results, we identify a set of coherent parameter settings that give particularly interesting simulation outputs in compliance with the model calibration objectives. The main contribution of the paper is to provide a re-useable method for calibrating a real-world ABM model.

## The SimpopLocal model

*A model to simulate the emergence of structured urban settlement system*
SimpopLocal is a stylized model describing an agrarian society in the Neolithic period, during the primary "urban transition" manifested by the appearance of the first cities (Schmitt, 2014). It is designed to study the emergence of a structured and hierarchical urban settlement system by simulating the growth dynamics of a system of settlements whose development remains hampered by strong environmental constraints. This exploratory model seeks to reproduce a particular structure of the Rank-Size distribution of settlements well defined in the literature as a generalized stylized-fact: for any given settlement system throughout the time and continents, the distribution of sizes is strongly differentiated, exhibiting a very large number of small settlements and a much smaller number of large settlements (Archaeomedes, 1998; Berry, 1964; Fletcher, 1986; Liu, 1996). This distribution can be modeled by a power-law when small settlements are not considered or a log-normal when small settlements are considered (Favaro, 2011; Robson, 1973). Such distributions are easily simulated by rather simple and non spatial statistical stochastic models (Gibrat, 1931; Simon, 1955). But for theoretical considerations and to specify the model in a variety of geographical frames, we think it necessary to make explicit the spatial interaction processes that generate the evolution of city systems. According to the evolutionary theory of system of cities (Pumain, 2009), the growth dynamics of each settlement is controlled by its ability to generate interurban interactions. The multi-agent system modeling framework enables us to include mechanisms, derived from this theory, that govern the interactions between settlements (Bretagnolle et al, 2006; Pumain and Sanders, 2013). The application of this concept resulted in several Simpop models (Bretagnolle and Pumain, 2010; Bura et al, 1996; Pumain et al, 2009; Sanders et al, 1997) in which the expected macro-structure of the log-normal distribution of sizes emerges from the differentiated settlement growth dynamics induced by the heterogeneous ability of interurban interactions. Therefore the aim of SimpopLocal is to simulate the hierarchy process *via* the explicit modeling of a growth distribution that is not entirely stochastic as in the Gibrat model (Gibrat, 1931) but that emerges from the spatial interactions between micro-level entities. Compared to the previous Simpop models, the originality of SimpopLocal is twofold. First, it is a simplified version which no longer qualitatively distinguishes the successive urban functions simulated during the evolution of urban systems but transposes them into an abstract innovation process having over time less and less impact in terms of wealth and population growth (this rule is in accordance with the concept of the decreasing efficiency of the improvement of an existing socio-economic system and prevents introducing any *a priori* amplification effect). Second, the appearance process of these innovations is an endogenous process that is linked to the size of the settlements.

*Agents, attributes, and mechanisms of the model*

The SimpopLocal model is a multi-agent model developed with the Scala programming language. It simulates the growth dynamics of agrarian settlements and their possible evolution toward urban settlements under strong environmental constraints that are progressively overcome by successive innovations.

The landscape of the simulation space is composed of hundred of settlements. Each settlement is considered as a fixed agent and is described by three attributes: the location of its permanent habitat, the size of its population, and the available resources in its local environment. The amount of available resources is quantified in units of inhabitants and can be understood as the carrying capacity of the local environment for sustaining a population which depends on the resources exploitation skills that the local population has acquired from inventing or acquiring innovation. Each new innovation acquired by a settlement develops its exploitation skills. This resource exploitation is done locally and sharing or trade is not explicitly represented in the model.

The growth dynamics of a settlement is modelled according to the assumption that its size is dependent on the amount of available resources in the local environment and is inspired from the Verhulst model (Verhulst, 1845) or logistic growth. For this experiment, we assume a continuous general growth trend for population - this may be different in another application of the model. The growth factor *r* is expressed on an annual basis, thus one iteration or step of the model simulates one year of demographic growth. The limiting factor of growth $R_M^i$ is the amount of available resource that depends on the number *M* of innovations the settlement *i* has acquired by the end of the simulation step *t*. $P_t^i$ is the population of the settlement *i* at the time *t* :

$$P_{t+1}^i = P_t^i \left[ 1 + r(1 - \frac{P_t^i}{R_M^i}) \right]$$

The acquisition of a new innovation by a settlement allows it to overtake its previous growth limitation by allowing a more efficient extraction of resources and thus a gain in population size sustainability. With the acquisition of innovations the amount of available resources tends to the maximal carrying capacity $R_{max}$ of the simulation environment:

$$R_M^i \xrightarrow{innovations\ acquisition} R_{max}$$

The mechanism of this impact follows the Ricardo model of diminishing returns (also a logistic model (Turchin, 2003)). The *InnovationImpact* represents the impact of the acquisition of an innovation and has a decreasing effect on the amount of available resources $R_{M+1}^i$ with the acquisition of innovations:

$$R_{M+1}^i = R_M^i \left[ 1 + InnovationImpact(1 - \frac{R_M^i}{R_{max}}) \right]$$

Acquisition of innovations can occur in two ways, either by the emergence of innovation within a settlement or by its diffusion through the settlement system. In both cases, interaction between people inside a settlement or between settlements is the driving force of the dynamics of the settlement system. It is a probabilistic mechanism, depending on the size of the settlement. Indeed, innovation scales super-linearly: the greater the number of innovations acquired, the bigger the settlement and the higher the probability of innovating (Arthur, 2009; Diamond, 1999; Lane et al, 2009). To model the super-linearity of the emergence of innovation within a settlement, we model its probability by a binomial law. If $P_{creation}$ is the probability that the interaction between two individuals of a same settlement is fruitful, *i.e.* leads to the creation of an innovation, and *N* the number of possible interactions, then by the Binomial law, the probability of the emergence of at least one innovation *P(m$_{creation}$>0)* can be calculated and then used in a random drawing :

$$P(m_{creation} > 0) = 1 - P(m_{creation} = 0)$$

$$= 1 - \left[\frac{N!}{0!(N-0)!} P_{creation}^{0} (1 - P_{creation})^{N-0}\right]$$

$$= 1 - (1 - P_{creation})^N$$

If the size of the settlement is $P_t^i$ then the number *N* of possible interactions between individuals of that settlement is :

$$N = \frac{P_t^i (P_t^i - 1)}{2}$$

The diffusion of an innovation between two settlements depends both on the size of populations and the distance between them. If *P$_{diffusion}$* is the probability that the interaction of two individuals of two different settlements is fruitful, *i.e.* leads to the transmission of the innovation, and K the number of possible interactions, then by the Binomial law the probability of diffusion of at least one innovation *P(m$_{diffusion}$ >0)* can be calculated and used in a random drawing :

$$P(m_{diffusion} > 0) = 1 - (1 - P_{diffusion})^K$$

But in this case, the size *K* of the total population interacting is a fraction of the population of the two settlements *i* and *j* which is decreasing by a factor *DistanceDecay* with the distance *D$_{ij}$* between the settlements, as in the gravity model (Wilson, 1971):

$$K = \frac{P_t^i \ P_t^j}{2 \ D_{ij}^{DistanceDecay}}$$

The process of population growth and the process of innovation creation/diffusion are re-iterated throughout the simulation. Because of the two positive feedbacks that operate on resource and population growth through the creation of innovation, the model is able to generate a very rapid expansion of settlements, an escalation of settlement growth. The simplest way to avoid situations where innovations are created in too large numbers which would lead to huge time consuming simulations, is to decide to stop the simulation when it reaches an arbitrary number of 10 000 innovations. In order to ensure the replicability of the model, the source code of SimpopLocal is filed in a public repository: http://iscpif.github.io/simpoplocal-epb/.

*Parameters to calibrate*
SimpopLocal has many parameters that have to be estimated in order to calibrate the model. Some can be empirically evaluated with the help of historical data and knowledge, while it is very difficult to give values to others. Those regarding the initial spatial distribution and organisation of settlements in the landscape can be approximated. The log-normal distribution of the settlement sizes and the central place theory (Christaller, 1933) for the geographical distribution of locations are models that are widely used by archaeologists to describe their spatial data (Archaeomedes, 1998; Johnson, 1977; Sanders, 2012), including Neolithic archaeological sites (Liu, 1996). In SimpopLocal, the mean density of that landscape and the average size of each settlement are representative of the usual orders of magnitude presented in these works. A hundred settlements are distributed according to these two theories and each settlement is initially composed of some 80 up to 400 inhabitants. Several scholars agree that an average annual growth of 0.02% is representative of the growth of agrarian settlements in the Neolithic times (Bairoch, 1985; Renfrew and Poston, 1979). The length of time required for a transition from agrarian to urban settlements is estimated according to (Bairoch, 1985; Marcus and Sabloff, 2008) to a couple of thousand years. We choose to operate our simulations on a four thousand year time period for settlements ranging from one hundred inhabitants up to about ten thousand inhabitants.

Because of a lack of empirical data, five parameters cannot be empirically approximated:

- $P_{creation}$, the probability that an innovation emerges from the interaction between two individuals of a same settlement.

- $P_{diffusion}$, the probability that an innovation is transmitted between two individuals of different settlements. We consider that the probability of diffusion is greater than the probability of creation, which means that copying is easier than inventing (Pennisi, 2010) in the model.

- *InnovationImpact*, the impact of the acquisition of innovation on the growth of settlements.

- *DistanceDecay*, the deterrent effect of distance on diffusion.

- $R_{max}$, the maximum carrying capacity of the landscape of each settlement (measured in number of inhabitants).

The model has been simplified as far as possible to retain only five parameters that cannot be empirically estimated. This number may seem rather low compared to the 40 or 50 parameters that were activated in the other versions of SIMPOP models, but this simplification allows a global exploration of the capabilities of SimpopLocal. Indeed, by means of intensive exploration, a calibrated state of the model can produce an estimation of the value of some parameters that could not be deduced from the empirical literature on archaeological settlements. These estimations will be useful afterwards for making predictions about the possible evolution of concrete settlements systems or comparing the evolutions of early urban systems in different regions. SimpopLocal is conceived for providing an open evolution that may lead to any type of size distribution of settlements depending on the parameter values. In order to estimate parameter values that could generate plausible settlement size distributions, we designed an automated calibration procedure.

## Designing an automatised calibration procedure
### Calibration as an optimisation problem
Model calibration is a procedure which seeks to minimize the difference between the behaviour simulated by the model and a behaviour defined according to expert knowledge and/or data. In most multi-agent systems the calibration is done manually by introducing values for the parameters and visually verifying that the output of the model corresponds to the expected results. But this method raises numerous problems highlighted by Stonedahl (Stonedahl, 2011): relationships between parameters are often non-linear; expertise during behaviour exploration may be biased by erroneous assumption; some parameters cannot be compared to empirical values; and manual exploration is tedious. In our first attempt to manually calibrate the model, we could not reach a stage where the calibration could reasonably be considered satisfying.

Because of the number of parameters and of the continuous scale of variation of their fields of variation, any exhaustive search strategy is not tractable either. First of all, this strategy generates a number of experiments that grows exponentially with the number of parameters under study. This would imply too gigantic an amount of computation in the case of SimpopLocal. Moreover, that strategy also produces a large quantity of data that then has to be processed and visualized. The search for patterns in the space of the dynamics of the model is thus replaced by the problem of a search for patterns in a database. Post-processing of such a large quantity of data is long and tiresome. Instead of resorting to exhaustive search methods, we choose to consider calibration as an optimisation exercise. The *a posteriori* exploration of results becomes then an *a priori* question: is there a parameter setting that would match our expectation?

Evolutionary algorithms have been established as a suitable solution to this way of considering calibration (Calvez and Hutzler, 2006; Stonedahl, 2011). They have been used to calibrate multi-agent systems in several application fields such as medicine (Castiglione et al, 2007), ecology (Duboz et al, 2010), economics (Espinosa, 2012; Stonedahl et al., 2010(a)), hydrology (Solomatine et al., 1999)... Despite the wide use of multi-agent systems in social sciences, this method has not yet been applied very often. To our knowledge only a few real world applications have been carried out (Heppenstall et al., 2007; Stonedahl et al., 2010(b)). Indeed this kind of numerical experiment remains a real challenge:

- it requires the definition of quantitative goals that evaluate the simulation outputs in a coherent manner which correspond to experts' expectations,

- it generates a massive computation load that requires technical skills and adapted infrastructures,

- it seeks to optimise noisy (stochastic) fitness function which is in itself a challenging exercise (Pietro et al., 2004).

To overcome these obstacles, cutting edge knowledge and tools in several fields of expertise have to be coupled within highly transdisciplinary teams involving social scientists, statisticians, optimization method specialists and distributed-computing experts.

*Exploring the parameter space given a set of objectives*
Evolutionary algorithms are heuristics that scan the search space using strategies inspired by natural processes to solve optimisation problems. In order to use evolutionary algorithms to calibrate a model, the first step is to formalize what the expected result of a "good" simulation is, which is, in our case, a suitable configuration of the settlement system. This suitable configuration reflects stylized facts that were established over the years by many scholars thanks to the exploration and processing of large amounts of empirical data. We have summarized this suitable configuration by three objective functions extracted as relevant stylised facts from the existing literature: a log-normal distribution of settlement sizes (Archaeomedes, 1998; Johnson, 1977; Liu, 1996; Sanders, 2012), a maximum size of ten thousands inhabitants (Bairoch, 1985; Marcus and Sabloff, 2008), and a four thousand years period duration for achieving this distribution (in any region of the world where agriculture was invented, the emergence of the first cities occurred a few thousand years later (Bairoch, 1985; Marcus and Sabloff, 2008).

The automatic technique of calibration that we propose is based on an evolutionary algorithm that explores the space of the parameter settings. This exploration is guided by these three objectives (Rey-Coyrehourcq, 2014; Schmitt, 2014). Each parameter setting (i.e. set of values for the 5 unknown parameters of the model) is evaluated according to the simulation outputs it produces. This evaluation measures the proximity between the outputs of simulation and the three objective functions defined for the model. This evaluation thus assesses the ability of the parameter settings to reproduce the stylized facts that we seek to simulate. The parameter settings receiving the best evaluations are then used as a basis for generating new parameter settings which will then be tested.

The SimpopLocal model being stochastic, the simulation outputs do vary for a given parameter setting from one simulation to the next. Therefore, the evaluation of the parameter setting on the three objectives must take into account this variability. We checked if 100 simulations for each set of parameters was sufficient to capture this variability and found out that it was a suitable compromise between capturing the variability and not increasing too much the computation duration. Here is how we test each set of parameters according to the three objective functions:

- The *objective of distribution* quantifies the ability of the model to produce settlement size distributions that fit a log-normal distribution. First we evaluate the outcome of each subset of 100 simulations corresponding to one parameter setting by computing, according to a 2-sample Kolmogorov-Smirnov test, the deviation between the simulated distribution and a theoretical log-normal distribution having the same mean and standard deviation. Two criteria are reported, with value 1 if the test is rejected and 0 otherwise: the likelihood of the distribution (the test returns 0 if p-value > 5%) and the distance between the two distributions (the test returns 0 if D-value < $D_\alpha^1$). In order to summarize those tests in a single quantified evaluation, we add the results of the two tests on the 100 simulations. The best possible score on this objective is thus 0 (all tests returning 0) and the worst 200 (all tests returning 1). By dividing this score by the worst possible value (200), we get a normalised error that facilitates the comparisons between parameter settings according to their scores on the three objectives.

- The *objective of population* quantifies the ability of the model to generate large settlements that have an expected size. The outcome of one simulation is tested by computing the deviation between the size of the largest settlement and the expected value of 10000 inhabitants: |(population of largest settlement - 10000)/10000|. The evaluation of the parameter setting reports the median of this test on the 100 simulations. This value represents the normalised error produced by the parameter setting being evaluated on the calibration (the closest to 0, the smallest the error). Computing this normalised error is not necessary but it facilitates the comparisons between parameter settings according to their scores on the three objectives.

- The *objective of simulation duration* quantifies the ability of the model to generate an expected configuration in a suitable length of time (in simulation steps). The duration of one simulation is tested by computing the deviation between the number of iterations of the simulation and the expected value of 4000 simulation steps: |(simulation duration - 4000)/4000|. The evaluation of the parameter setting reports the median of this test on the 100 simulations. This value represents the normalised error produced by the parameter setting being evaluated on the calibration. This normalised error (the closest to 0, the smallest the error) facilitates the comparisons between parameter settings according to their scores on the three objectives.

---

[1] Computed with α = 1.36.

The three objective functions constitute a fitness function of our optimisation problem which is therefore multi-objective. To solve such problems, it is sometimes possible to aggregate the vector values in a single scalar value reflecting an absolute quality of the solution. In our case it is not possible to provide a meaningful aggregated function. Thus we rely on a multi-objective optimisation algorithm. This type of algorithm computes compromise solutions: among those solutions none dominates the others by presenting better scores on all three objectives at the same time. Such solutions are called *Pareto compromise* and the entire set of Pareto compromise solutions is called the *Pareto front*.

### *Executing a large scale distributed evolutionary algorithm with a noisy time-consuming fitness function*

The use of global methods of search like evolutionary algorithms for the calibration of a multi-agent model (and especially a stochastic multi-agent model) entails a high computational cost (Sharma et al, 2006). This kind of load is too large to be executed on local computers. Super-computers are very expensive and not easily available in most laboratories, therefore we decided to develop a calibration framework that could run on both expensive super-computers and on computing grids which make computing power more accessible by federating it on a world wide scale. For the experiment presented in this paper we used the European Grid Infrastructure (EGI).

Computing grids offer a solution for the resolution of such computationally intensive problems, however computing at such a large scale is a true challenge. It supposes orchestrating the execution of tens of thousands of instances of the model on computers distributed all over the world. The cumulated probability of local breakdowns and the impossibility to optimally distribute the workload on the grid system makes its effective use very difficult. To overcome these difficulties we have used the OpenMOLE platform (Open Model Exploration, www.openmole.org) (Reuillon et al, 2010, 2013), which provides a bridge that helps modellers to cross the technical and methodological gap which separates them from high performance computing. OpenMOLE is a dedicated, textual and graphic language, exposing coherent bricks at the right level of abstraction to conceive reusable experiments in order to solve inverse problems on models. The methods of resolution are described independently of a particular model and thus support the reproducibility and the re-use of the experimental numerical methods between the modellers.

For calibrating SimpopLocal we implemented a classical multi-objective evolutionary algorithm called SMS-EMOA (Emmerich 2005) in a distributed manner, on top of OpenMOLE. The parameters of this algorithm are exposed in table 1. The code of this workflow and the code for the fitness computation are available on: http://iscpif.github.io/simpoplocal-epb/.

**Table 1. Parameters of the implemented evolutionary algorithm**

| Evolutionary Algorithm | SMS-EMOA |
|---|---|

| Genome | $P_{creation}$ [0 ; 1], $P_{diffusion}$ [0 ; 1], InnovationImpact [0 ; 2], DistanceDecay [0 ; 4], $R_{max}$ [1 ; 40 000] |
|---|---|
| Objective Function | Multi-objective : Distribution, population and simulation duration |
| Crossover | SBX RGA operator with Bounded Variable modification, *see APPENDIX A of (Deb K, 2000)* -> with a distribution index equals to 2.0 and a crossover rate equals to 0.5 |
| Mutation | Gaussian mutation and self adaptation (Hinterding R, 1995) |
| Dominance | Epsilon dominance, -> epsilons = distribution 0.0, population 10.0, simulation duration 10.0 |
| Nadir | distribution 500, population 100000, simulation duration 10000 |
| Selection | Binary tournament |
| Population | 200 |

This algorithm evaluates millions of parameter settings to approximate the Pareto front. In its most efficient implementation one execution of SimpopLocal lasts 1.5 seconds on a current processor. The automatic calibration of SimpopLocal requires the simulation of almost 500 million executions of the model, which would represent nearly 20 consecutive years on a single computer. To achieve this huge computation load, the SMS-EMOA algorithm was distributed on the numerous computers of the EGI by using the technique known as the "island model" (Emmerich 2005; Whitley et al, 1997). The classical island models consist in instantiating permanent islands (isolated instances of an evolutionary algorithm) on many computers and organising migration of solutions between those islands. The EGI grid is a worldwide batch system on which organising direct communications between islands running on multiple execution nodes is very challenging. Thus the classical island model has been adapted to the EGI architecture.

During the computation, a central population of 200 parameter settings (here called individuals) is maintained on the computer that orchestrates the submission of the computing jobs on the grid. Each job computes the evolution of the population of an island, which is an independent instance of SMS-EMOA seeded with 50 individuals randomly sampled among the central population of 200 individuals. The "island job" life cycle is managed by the EGI : each job is submitted to the EGI, then to the queue of a cluster and finally starts running when a slot becomes available on the cluster. When it starts running it is configured to run for 1 hour. 5000 concurrent jobs are maintained on the grid at any time.

Once an island job is finished its final population of individuals (i.e. parameter settings) is transferred back to the submission computer and merged into the global population using the elitism algorithm of the SMS-EMOA (based on the contribution to the hypervolume of the Pareto). A new island job is then submitted to the grid. This algorithm is run until 190 000 island jobs have been completed.

As previously stated, the fitness function of this experiment is stochastic, so we execute the model a hundred times for each parameter setting evaluation. Despite this precaution, it is well-known that among the millions of executions computed by the algorithms, some solutions might get "lucky" and be over-estimated, creating a bias in the evolution process (Pietro et al, 2004). According to Pietro et al. (2004) who reviewed various method to tackle this problem, we choose to re-evaluate already evaluated solutions on a regular basis to prevent "*from retaining incorrect fitness*" (p.2). A specific strategy was implemented: one over a hundred parameter setting evaluations was used to re-evaluate already evaluated parameter settings instead of evaluating a new parameter setting. A parameter setting which has been over-estimated the first time and thus kept among the optimal individuals by mistake, will probably be correctly evaluated the second time and eliminated from the optimal selection.

## Results and discussion

Evolutionary algorithms are good solutions when no exhaustive method is available, however they don't ensure the optimality of the computed solution. Because of their heuristic nature, they are procedures that can run indefinitely. In this work we have used the number of evaluated islands as a stopping criterion given the computing power that was at our disposal. That way we preconditioned the experience so that the total execution would last for 3 to 4 days, which means that the evolutionary optimisation algorithm was stopped after the execution of 190000 islands, which represents about 200000 computation hours (200000 CPU hours[2]). An evolutionary algorithm is declared "converged" when it makes no further improvements in the search for good solutions. One of the best metrics presently available to measure the convergence of multi-objective optimisation algorithm is the stagnation of the *hypervolume* (Fonseca et al, 2006; Naujoks et al, 2005; Zitzler and Thiele, 1998). The hypervolume measures the volume of the dominated portion of the objective space and its stagnation indicates that the algorithm has converged. Figure 1 depicts the evolution of the hypervolume in function of the number of evaluated islands, which can also be quantified in terms of computation time or CPU hours. After the execution of 120000 CPU hours, the hypervolume stabilises. This stability in the long term (70000 CPU hours) indicates that the algorithm has converged. The Pareto front probably wouldn't have improved much more with further computation time. This result shows that despite the stopping criterion that we have chosen because of technical limitations, the proposed parameter settings of the Pareto front are seemingly the best possible results that can be obtained with this method and they most likely correspond to the global optimum.

---

[2] Central Processing Unit hours

*Figure 1 : Hypervolume enclosed by the Pareto front and the converging evolution of the evolutionary algorithm under the constraint of 3 objectives[3].*

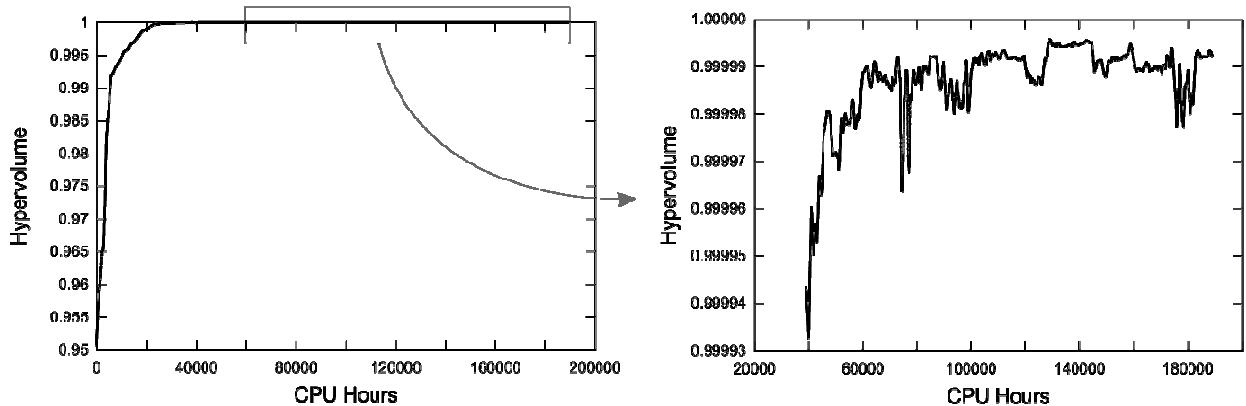

After the convergence of the algorithm, 200 different parameter settings were proposed as possible calibrated sets of parameter values. Because a set of 100 executions can only provide an approximation of the objective scores of a parameter setting and because we want to make sure that they are well evaluated, 10000 more executions of each proposed parameter settings where conducted. After re-computing the dominance selection, the dominated parameter settings were excluded and 62 parameter settings remained in the Pareto front. As said above, a multi-objective exploration does not lead to the selection of a single optimised solution but leads to a set of possible candidates for the calibration: each set of parameter settings selected by the procedure represents a specific compromise on the three objectives. It is thus possible to distinguish among the parameter settings those satisfying only two objectives out of three from those which offer a better compromise on the three objectives without however reaching the best possible values on each objective. This configuration confirms that these three objectives are not redundant and that the procedure vacillates between them to find well-fitted parameter settings. Among the 62 parameter settings of the Pareto front, some are less satisfying than others. We consider as unsuited the parameter setting that have at least one of the scores of the three objectives that has a normalised error over 0.1, i.e. over 10% of error. Only 29 parameter settings satisfy this new condition. The values estimated for each parameter of the 29 selected sets are presented in table 2.

**Table 2. Calibrated Parameter settings[4]**

| PARAMETER SETTING | | | | | % ERROR ON EACH OBJECTIVE SCORE | | |
|---|---|---|---|---|---|---|---|
| $R_{max}$ | Distance-Decay | $P_{creation}$ | $P_{diffusion}$ | Innovation-Impact | %error Distribution-Objective | %error Population-Objective | %error Time-Objective |
| 10464 | 0.691 | 1.11E-06 | 7.90E-07 | 0.0077 | 2.02 | 6.59 | 1.00 |
| 10465 | 0.709 | 1.14E-06 | 7.88E-07 | 0.0078 | 2.04 | 7.31 | 0.10 |
| 10459 | 0.705 | 1.15E-06 | 7.92E-07 | 0.0077 | 2.22 | 6.98 | 0.48 |
| 10465 | 0.708 | 1.15E-06 | 7.86E-07 | 0.0078 | 2.51 | 6.00 | 0.40 |

---

[3] In order to smooth the local variability of the raw hypervolume value, it has been averaged using a 1000 island sliding window.
[4] For readability the values presented are rounded.

| | | | | | | | |
|---|---|---|---|---|---|---|---|
| 10261 | 0.679 | 1.17E-06 | 7.78E-07 | 0.0078 | 2.79 | 5.29 | 4.55 |
| 10262 | 0.679 | 1.15E-06 | 7.52E-07 | 0.0078 | 2.95 | 5.00 | 2.15 |
| 10262 | 0.665 | 1.14E-06 | 7.24E-07 | 0.0078 | 2.99 | 5.53 | 1.23 |
| 10261 | 0.683 | 1.12E-06 | 7.38E-07 | 0.0080 | 3.1 | 4.34 | 1.15 |
| 10260 | 0.699 | 1.17E-06 | 7.38E-07 | 0.0079 | 3.62 | 5.51 | 0.20 |
| 10287 | 0.690 | 1.23E-06 | 7.56E-07 | 0.0078 | 3.63 | 3.74 | 3.46 |
| **10259** | **0.688** | **1.20E-06** | **7.41E-07** | **0.0079** | **3.74** | **3.55** | **2.48** |
| 10169 | 0.736 | 1.29E-06 | 7.39E-07 | 0.0079 | 4.90 | 5.20 | 0.03 |
| 10205 | 0.683 | 1.19E-06 | 7.42E-07 | 0.0082 | 5.10 | 2.60 | 6.03 |
| 10126 | 0.738 | 1.22E-06 | 7.61E-07 | 0.0082 | 5.69 | 2.88 | 1.50 |
| 10126 | 0.738 | 1.24E-06 | 7.39E-07 | 0.0082 | 6.02 | 3.08 | 0.55 |
| 10096 | 0.701 | 1.14E-06 | 7.14E-07 | 0.0084 | 6.12 | 2.58 | 1.55 |
| 10169 | 0.736 | 1.29E-06 | 7.39E-07 | 0.0080 | 6.25 | 2.46 | 1.20 |
| 10165 | 0.734 | 1.29E-06 | 7.24E-07 | 0.0080 | 6.31 | 2.91 | 0.30 |
| 10121 | 0.732 | 1.28E-06 | 7.41E-07 | 0.0081 | 6.41 | 2.36 | 1.90 |
| 10164 | 0.735 | 1.29E-06 | 7.27E-07 | 0.0080 | 6.45 | 2.74 | 0.45 |
| 10103 | 0.733 | 1.24E-06 | 7.42E-07 | 0.0084 | 7.67 | 1.90 | 3.10 |
| 10092 | 0.736 | 1.29E-06 | 7.14E-07 | 0.0082 | 7.81 | 2.22 | 1.10 |
| 10098 | 0.737 | 1.29E-06 | 7.12E-07 | 0.0082 | 7.84 | 2.55 | 0.58 |
| 10094 | 0.741 | 1.28E-06 | 7.12E-07 | 0.0083 | 8.46 | 1.99 | 1.00 |
| 10129 | 0.737 | 1.29E-06 | 7.07E-07 | 0.0082 | 8.64 | 1.97 | 0.68 |
| 10110 | 0.735 | 1.28E-06 | 6.77E-07 | 0.0083 | 9.04 | 2.48 | 0.03 |
| 10091 | 0.744 | 1.31E-06 | 7.25E-07 | 0.0083 | 9.22 | 1.51 | 2.68 |
| 10091 | 0.741 | 1.31E-06 | 7.12E-07 | 0.0083 | 9.61 | 1.49 | 2.15 |
| 10109 | 0.734 | 1.28E-06 | 6.79E-07 | 0.0084 | 9.64 | 1.77 | 0.18 |

The analysis of this new subset shows interesting aspects. First, all of them correctly identify the order of magnitude of the only parameter of the model which can be roughly deduced from the initial conditions and the calibration objectives: the maximum resource parameter ($R_{max}$) which is used to limit the logistic growth of the population of each settlement[5]. Second, the four parameters whose values are *a priori* unpredictable (*DistanceDecay*, the parameter of dissuasion from the interactions by the distance; $P_{creation}$, the probability of appearance of an innovation; $P_{diffusion}$, the probability of its diffusion; and finally *InnovationImpact*, the impact of the innovation on the growth of the population) are estimated in a very small domain of variation compared with their possible domain of variation (table 3). The ratio between the estimated and theoretical volume defined by these five dimension domains is of ~ $4.7^{-16}$/ 320000 = $1.5^{-21}$ ! What is remarkable is how the values taken by each parameter are all in a small neighbourhood, which suggests high reliability considering that they must be given those orders of magnitude to obtain plausible results while using the model for simulation. However it should be noted that the exact values estimated for each parameter do not have any absolute meaning. They only make sense all together, according to their interrelationships in the mechanisms of the model.

---

[5] Non-linear effects make this estimation tricky: the expected value would have been 10000 inhabitants, according to the objective of population, but in order to reach this value, the maximum carrying capacity of the landscape must be slightly higher.

**Table 3. Global domain of exploration and calibrated parameter value domain of variation**

|  | **Explored domain of variation** | **Domain enclosed in the Pareto front** |
|---|---|---|
| $P_{creation}$ | [0 ; 1] | ~[ 1.1E-06; 1.3E-06 ] |
| $P_{diffusion}$ | [0 ; 1] | ~[ 6.7E-07; 6.9E-07 ] |
| InnovationImpact | [0 ; 2] | ~[ 7.7E-03; 8.4E-03 ] |
| DistanceDecay | [0 ; 4] | ~[ 0.66; 0.75 ] |
| $R_{max}$ | [1 ; 40 000] | ~[ 10090; 10465 ] |
| ***Domain Volume*** | ***320000*** | ***~ 4.7E-16*** |

Within the 29 parameter settings subset there is no way to prefer a setting over the others: the multiple executions of each setting lead to good results and acceptable variability of the outputs. Figure 2 and 3 give an example of such outputs. They depict the results produced by one of the parameter settings (in bold characters[6] in table 2). Figure 2 shows the evolution of the rank-size distribution of settlement sizes during a simulation of the parameter setting. This evolution corresponds to what is expected of the model: a progressive and continuous process of hierarchical organisation of the settlement system (the slope of the linear fit of the rank-size distribution shifts from 0.2 to 0.9 in 4000 years for a maximum size reached of about 10000 inhabitants). This result is quite robust if we consider the low variability of the recorded final state from one simulation to another (figure 3).

**Figure 2: Evolution of the Rank-Size distribution during a simulation of one of the best calibrated parameter settings.**

---

[6] The non-rounded parameter setting is : {$R_{max}$ : 10259, 331894632433; DistanceDecay : 0, 6882107473716844; $P_{creation}$ : 1.2022185310640896E-06; $P_{diffusion}$ : 7.405303653131592E-07; InnovationImpact : 0.007879556611500305}. The seed used for the simulation depicted on figure2 is: −6863419716327549772.

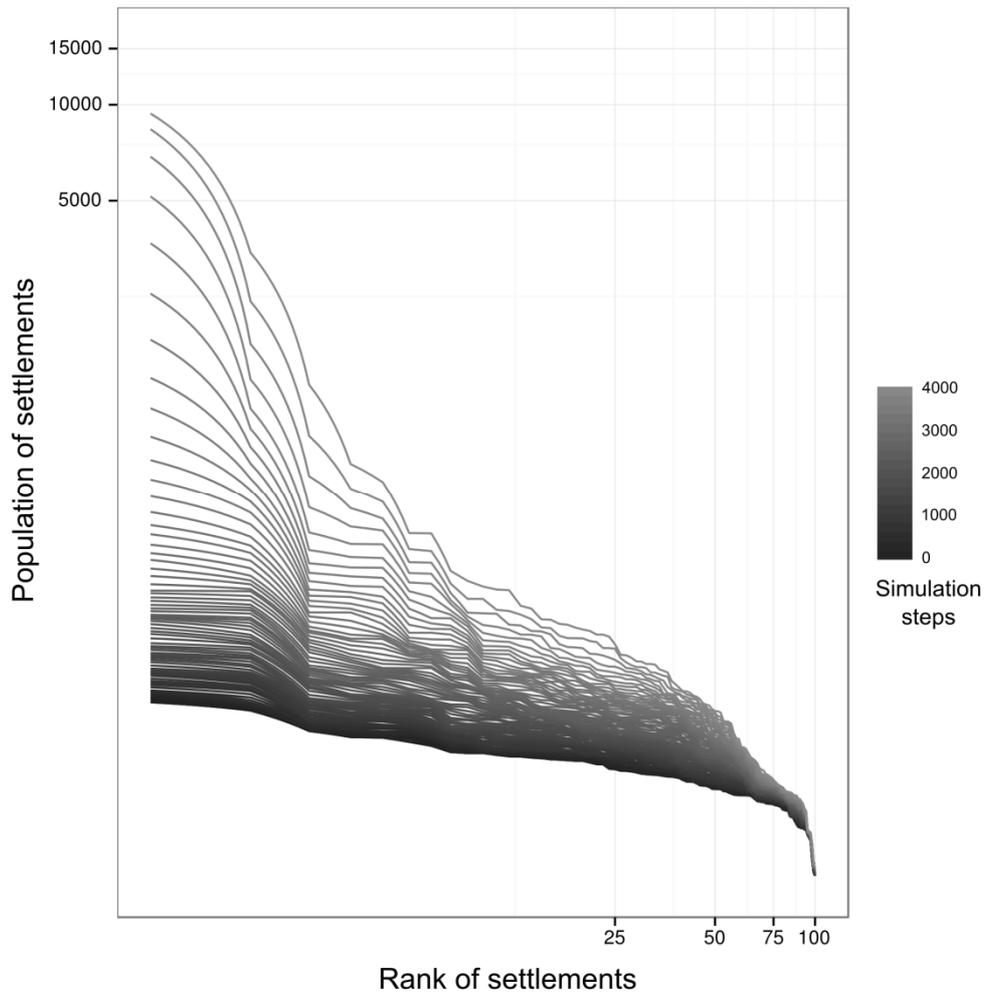

**Figure 3: Variability of the Rank-Size distribution at the final simulation step on 100 simulations of one of the best calibrated parameter settings.**

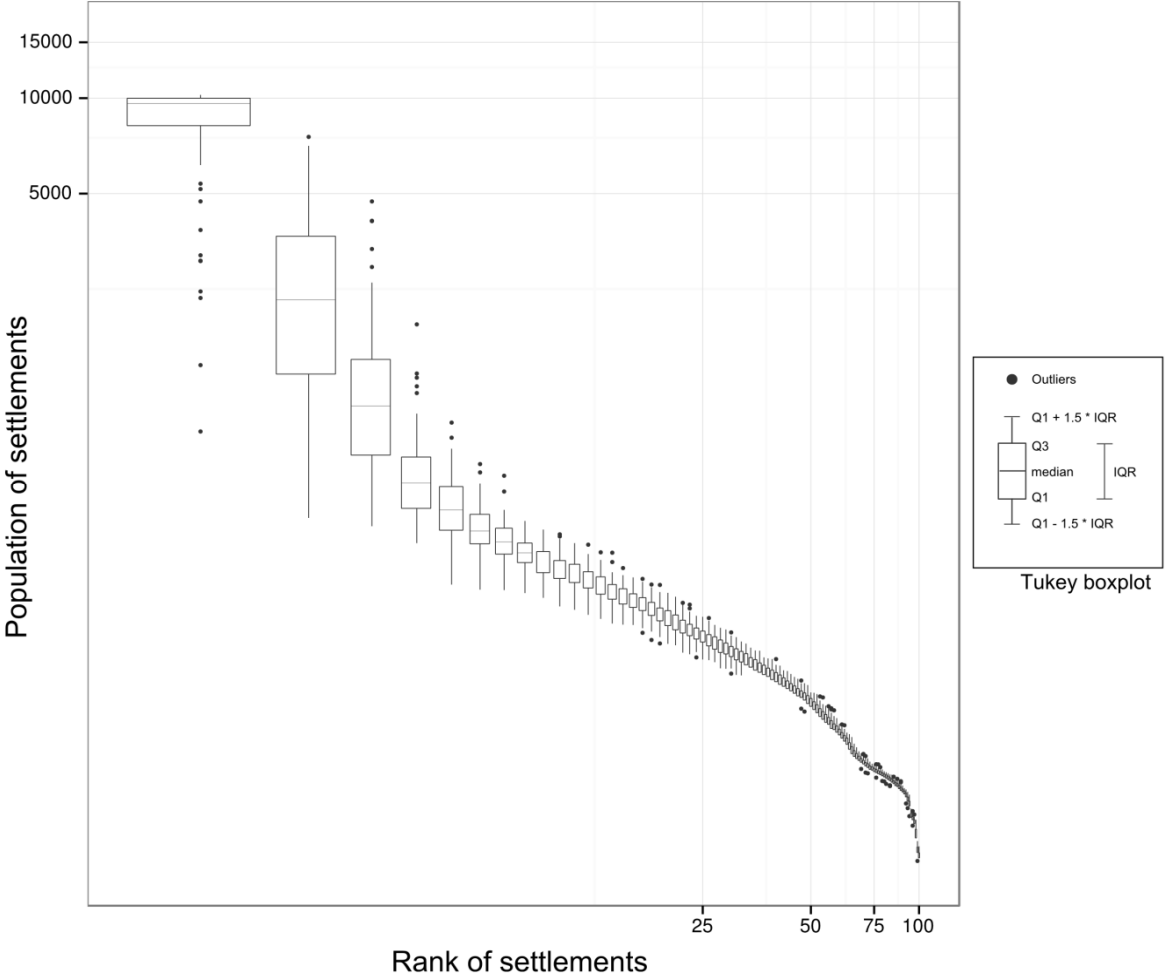

## Conclusion

The automatic calibration procedure used in this article was applied for the first time on a multi-agent model intended to simulate the emergence of a system of urban settlements called SimpopLocal. It provides convincing results to solve the calibration problem for multi-agents models. With five parameters whose value could not be estimated from empirical data, it would have been quite difficult to find an estimation of a calibrated parameter setting "manually". By reversing this calibration problem into an optimisation problem, this automated calibration procedure generates parameter settings which reproduce the stylized facts embodied in three objective functions very well. This enables us to confirm a decisive advance in the validation of this model : we proved that the SimpopLocal model, as conceived in its simplicity, is able to satisfactorily generate a plausible evolution including the emergence of an urban hierarchy.

This paper illustrates the potential of a new modelling procedure for exploring the dynamics of geographical systems which may be useful as well for other social sciences. The novelty is in the combination of three principles: a conceptual framework making explicit and quantifying the criteria for evaluating the model; a methodology for exploring the parameter space with an automated process using evolutionary algorithms; a technical protocol for reducing the exploration time by massive distributed computing. The application of these principles to modelling not only improves the quality of communication about the model, but as well ensures the repeatability of its experimentation and contributes to create accessible modelling tools that can be shared via the free and open source software tool for model experiments, OpenMOLE (*www.openmole.org)*. In order to help the dissemination of these techniques, we published several tutorials on the developed tools : on how to design a grid exploration of a NetLogo model[7], how to explore a NetLogo model with evolutionary algorithms[8], etc. This work forges a path for other modelers, who could use a similar data-intensive grid-based approach in their modeling experiments.

---

[7] http://www.openmole.org/documentation/tutorials/netlogo-on-the-grid/
[8] http://www.openmole.org/documentation/tutorials/ea-with-netlogo/